\documentclass{iopart}

\usepackage{graphicx,comment}
\usepackage{dcolumn}
\usepackage{bm}
\usepackage{amssymb,amsfonts}
\usepackage[english]{babel}
\usepackage{geometry}
\usepackage{braket}
\usepackage{float}
\usepackage{caption}
\usepackage{subcaption}
\usepackage[dvipsnames]{xcolor}
\usepackage{multirow}
\usepackage{appendix}
\usepackage{hyperref}

\newcommand{\newblock}{\ }

\begin{document}

\title{Trends in hot carrier distribution for disordered noble-transition metal alloys}

\author{Eklavya Thareja$^1$, Kevin M. McPeak$^2$, Phillip T. Sprunger$^1$, Ilya Vekhter$^1$, William A. Shelton$^1$}
\address{$^1$Department of Physics and Astronomy, Louisiana State University, Baton Rouge, Louisiana 70803, USA}
\address{$^2$Cain Department of Chemical Engineering, Louisiana State University, Baton Rouge, Louisiana 70803, USA}

\ead{\mailto{eklavya.thareja@gmail.com}}

\date{\today}

\begin{abstract}
We developed and tested an approach for predicting trends for efficient hot carrier generation among disordered metal alloys. We provide a simple argument for the importance of indirect transitions in the presence of disorder, thus justifying the use of Joint Density of States ($JDOS$)-like quantities for exploring these trends. We introduce a new $JDOS$-like quantity, $JDOS_K$, which heuristically accounts for longer lifetimes of quasiparticles close to the Fermi energy. 
To demonstrate the efficacy of this new quantity, we apply it to the study of Cu$_{50}$X$_{50}$ where X = Ag, Au, Pd and Y$_{50}$Pd$_{50}$ where Y = Au, Ni. We predict that Ni$_{50}$Pd$_{50}$ produces the most hot carriers among the alloys considered. The improvement in the density of excited photocarriers over the base alloy used, Cu$_{50}$Ag$_{50}$, is 3.4 times for 800 nm and 19 times for 1550 nm light. This boost in hot-carrier generation is consequence of the ferromagnetic nature of the Ni alloy. We argue that our method allows efficient material-specific predictions for low bias photoconductivity of alloys.
\end{abstract}

%
%
%
%
%

\section{Introduction}
{
Generating "hot" carriers above equilibrium in metals holds potential for efficiently converting photons into electrical charge carriers in near-infrared (NIR) optoelectronics. Yet, identifying metals capable of both near-infrared hot-carrier generation and sustaining carrier lifetimes proves challenging. Disordered alloys frequently exhibit unique behaviors not found in their constituent elemental compounds, providing novel properties and alternative design approaches compared to ordered metals. Alloying can boost hot carrier generation via (i) disorder broadening of the states that can lead to increased DOS near photon energy below the Fermi level relative to an ordered systems, and 
(ii) increase in number of excited electrons due to disorder-enabled indirect transitions. Thus, alloying provides more phase space for excitations 
relative to ordered systems that  are usually driven by direct (momentum-conserving) transitions in $\vec{k}$-space. Several of us recently showed that noble-transition metal, Au$_{x}$Pd$_{1-x}$, produces more hot carriers with longer lifetimes than its constituent metals with the near-infrared light \cite{AuPd_AM}. Alloying can even be used to control hot carrier relaxation time \cite{Memarzadeh_2020} and improve absorption in thin films \cite{Munday_2020}. 
However, both the number of elements and compositional space of alloys is extremely large. This makes searches for potentially improved hot carrier materials immensely difficult, and a reliable method for predicting and comparing trends associated with 
hot-carrier generation for different alloys has been lacking.  Having an approach that can rank candidates and also provide additional insight into why one system is better than another is needed.}

{Joint density of states ($JDOS$), which is a convolution of the energy distribution of filled and empty states that differ by photon energy used for excitation, has been used for comparing candidate disordered alloys for hot carrier generation~\cite{Munday_2015}. Here we first provide justification for using $JDOS$-like quantities as the figure of merit, by explicitly showing that in the presence of disorder the contributions of directly and indirectly excited carriers to the conductivity are of the same order except for very dilute alloys. This justifies going from the momentum conserving transitions to convolution of the densities of states.  However, we go further by noting that electrons excited into states away from the Fermi surface generically have shorter lifetimes than the low-energy excitations, in part 
due to inelastic scattering. Thus, it is the low energy photoexcited carriers that contribute more to low bias photoconduction. To account for this we define a new figure of merit,  $JDOS_{K}$, which introduces into the convolution of the densities of states at energies separated by the photon energy a kernel, $K$, that gives larger weight to the 
excited electrons near the Fermi surface. 
We elucidate the differences between $JDOS$ and $JDOS_{K}$, and apply our approach to investigating hot carrier generation trends in the noble-transition metal alloys
Cu$_{50}$Ag$_{50}$, Cu$_{50}$Au$_{50}$, Cu$_{50}$Pd$_{50}$, Au$_{50}$Pd$_{50}$ and Ni$_{50}$Pd$_{50}$.}



{The activation of hot carriers through photoexcitation in metals holds significance across various physical phenomena, encompassing surface photochemistry \cite{Heinz_93}, photocatalysis \cite{Halas_2022}, magneto-optical recording \cite{Ogawa_97}, and the transport of electrical charge and heat \cite{Ogawa_97}. Metals, when photoexcited, can produce hot carriers through interband (such as $d \rightarrow sp$ or $d \rightarrow d$), intraband ($sp \rightarrow sp$) excitations, and Landau damping assisted by plasmons~\cite{Pawlik_97,Spicer_64,Fatti_2000,Atwater2014,Khurgin_2015,Hoang2017}. 
Although nanostructured plasmonic materials enable adjustable distribution of hot-carrier energy, the efficiency of resulting devices remains notably low \cite{Spicer_64}. Leveraging the excitation of abundant $d$-band states presents a pathway to enhance the efficiency of hot carrier devices \cite{Fatti_2000}. Consequently, materials featuring low-lying $d$-band states could facilitate the translation of these enhancements to near-infrared (NIR) hot carrier applications, such as ultrafast NIR photodetection.}



Previously, Stofela et. al. \cite{AuPd_AM} and Manoukian et. al. \cite{CuPd_aip} have synthesized and characterized Au$_{50}$Pd$_{50}$ and Cu$_{1-x}$Pd$_{x}$ noble-transition metal alloy films.  When the Au$_{50}$Pd$_{50}$ system was excited at 1550 nm, it displayed a 20-fold increase in hot holes compared to Au films and three-times longer lifetimes than pure Pd. Unlike Au,  Cu does not have a strong spin-orbit interaction, however, our previous study did find common band behavior between the Cu 3-$d$ and Pd 4-$d$ states. 
{This phenomenon is attributable to robust $d$-band hybridization as indicated by  photoelectron spectroscopy and density functional theory calculations \cite{AuPd_AM,CuPd_aip}. Furthermore, ultrafast spectroscopy demonstrates composition dependent post-excitation dynamics, where dilute Pd alloys exhibit non-thermalized hot carriers right after excitation. The NIR inter-band transitions observed in these Pd alloys offer a pathway to non-thermalized hot holes, which are integral to several proposed optoelectronic applications \cite{Besteiro2021,Reddy2021,Brongersma2015}.}

This raises the question whether better optoelectronic devices can be made using disordered alloys of 3$d$-transition metal and noble metals. 
{Ordered alloys and disorder using special quasi-random structure method have been studied \cite{Matej_2024}. However, for disordered alloys broadening will substantially affect optical transitions, thus performing configurational average is crucial.} 
Hence, we use KKR-CPA (Korringa, Kohn, and Rostoker method with Coherent-Potential Approximation) to perform these calculations~\cite{Stocks_94,Stocks_92,Soven_67,Korringa_47,Kohn_54}, see section 3 for details.
We use aforementioned scheme involving $JDOS_K$ to study disordered noble-transition metal alloys. We focus on FCC binary alloys and fix the composition to 50-50 to emphasize trends and establish validity of our approach. We find that amongst the alloys considered Ni$_{50}$Pd$_{50}$ will generate the most hot carrier for low-bias photoconduction using 800 nm and 1550 nm light.

{Remainder of the paper is organized as follows. In section 2, we discuss importance of indirect transition in disordered alloys and we introduce a new figure of merit, $JDOS_K$, to use for low bias photoconductance. We discuss computational details of our calculation in section 3. In section 4, we use $JDOS_K$ to rank various noble-transition metal disordered alloys. In section 5, we summarize our findings.}


\section{Indirect Transitions and Figure of Merit \label{sec:indirect}}

In pure materials, optical transitions conserve the electron momentum, and, therefore, numerical evaluation of their amplitude requires a fine momentum mesh in the energy window comparable to the photon energy. Disorder relaxes the momentum conservation requirements, allowing for the indirect transitions, and, as we show now, in sufficiently disordered alloys the $JDOS$ provides a good figure of merit for the probability of optical transitions, and thus hot carrier generation. 

One can expect this result on general grounds for a random alloy. Metallic bonding assumes strong sharing of the electrons in atomic orbitals on neighboring lattice sites. The dipole selection rules dictate that only states of opposite parity can be connected by the optical transitions in pure systems. However, random substitutional disorder mixes states of different parity since locally the inversion symmetry is broken. One may argue that this effect is strong for substitutions with ions of different size, since the lattice distortions decay algebraically, as power law in distance from the impurity site, and therefore the inversion symmetry breaking is persistent at long distances even for a dilute concentration. That symmetry breaking enables previously forbidden transitions. Averaging over random local environments is equivalent to allowing transitions between any states respecting energy conservation, even if momentum is not conserved. For the random alloys close to 50-50 concentrations, such as those studied here, $JDOS$ should provide a qualitatively, or even semi-quantitatively, correct estimate of the transition probability.

This argument can be put on even stronger footing for alloys with different ionic charges of constituents, considering the effects of the Coulomb potential. Here we make an argument similar to the one outlined by Ziman \cite{Ziman1960}. For simplicity, and without loss of generality, we consider single band system with the states labeled by the crystal momentum, $|\bm k\rangle$. The perturbative correction to the band Hamiltonian due to impurities and interaction with photons is
\begin{equation}
	H_{pert} = (e/m)\bm{p}\cdot\bm{A} + V_{imp}\,, \qquad V_{imp}=\sum_{i=1}^{N_{imp}} V_0 (\bm r- \bm R_i)\,,
\end{equation}
where $\bm{p}$ is the momentum operator, $\bm A$ is the vector potential of the electromagnetic field, and $V_0$ is the  scattering potential due to each of $N_{imp}$ impurities located at positions $\bm R_i$. Let us consider the optically induced transition from the initial state, $\bm k_i$ to the final state, $\bm k_j$. The two lowest order contributions to the linear, in $\bm A$, response are the direct transitions appearing at the first order,
	$\braket{\bm k_j|\bm{p}\cdot\bm{A}|\bm k_i}$,
and impurity-enabled indirect transitions appearing at the second order 
\begin{equation}
 \sum_{m}\frac{\braket{\bm k_j|\bm{p}\cdot\bm{A}|\bm k_m}\braket{\bm k_m|V_{imp}|\bm k_i}}{E_i- E_j}\,.
 \label{eq:indirect}
\end{equation} 
For charged impurities the screened potential $V_0(r)= (Ze^2/r) e^{-q_{TF}r}$, where $Z$ is the impurity charge, and the Thomas Fermi screening wave vector in the free electron gas approximation is $q_{TF}^2 = 4\pi e^2 n(E_F)$, with $n(E_F)$ as the density of states at the Fermi energy. For most metals the screening length is of the order of lattice spacing, and therefore $q_{TF}$ is comparable to the size of the Brillouin Zone. In the momentum  representation
\begin{equation}
	V_{0}(q)= \frac{4\pi Ze^2}{q^2 + q_{TF}^2}\approx \frac{4\pi Ze^2}{q_{TF}^2}\approx \frac{Z}{n(E_F)}\,,
\end{equation}
over most of the Brillouin Zone where we can take $q<<q_{TF}$. 
Independence of this approximate value on the momentum transfer suggests that, to a good approximation, the matrix element $\braket{\bm k_m|V_{imp}|\bm k_i}\approx N_{imp} \braket{\bm k_m|V_0|\bm k_i}$. For a rough estimate we take the typical energy difference $E_i - E_j \sim E_F$, and note that 
for moderately high concentrations of impurities (as in our 50-50 alloys), the number of impurities is of the order of the number of electrons, $N_{imp}\approx N$. In that case the contribution of the indirect transitions in Eq.~\ref{eq:indirect} is of order
\begin{equation}
    \frac{\braket{\bm k_j|\bm{p}\cdot\bm{A}|\bm k_m}\braket{\bm k_m|V_{imp}|\bm k_i}}{E_i- E_j}\simeq
    \braket{\bm k_j|\bm{p}\cdot\bm{A}|\bm k_m}\frac{N}{E_F n(E_F)}\,.
\end{equation}
The latter ratio is of order unity since $n(E_F)\sim N/E_F$. We therefore conclude that in disordered alloys the indirect optical transitions are not suppressed. The sum over all the intermediate states $|\bm k_m\rangle$ in Eq.~\ref{eq:indirect} now implies that we can connect all possible initial and final states separated by the photon energy.

\begin{figure}[t]
\centering
\includegraphics[width=\columnwidth]{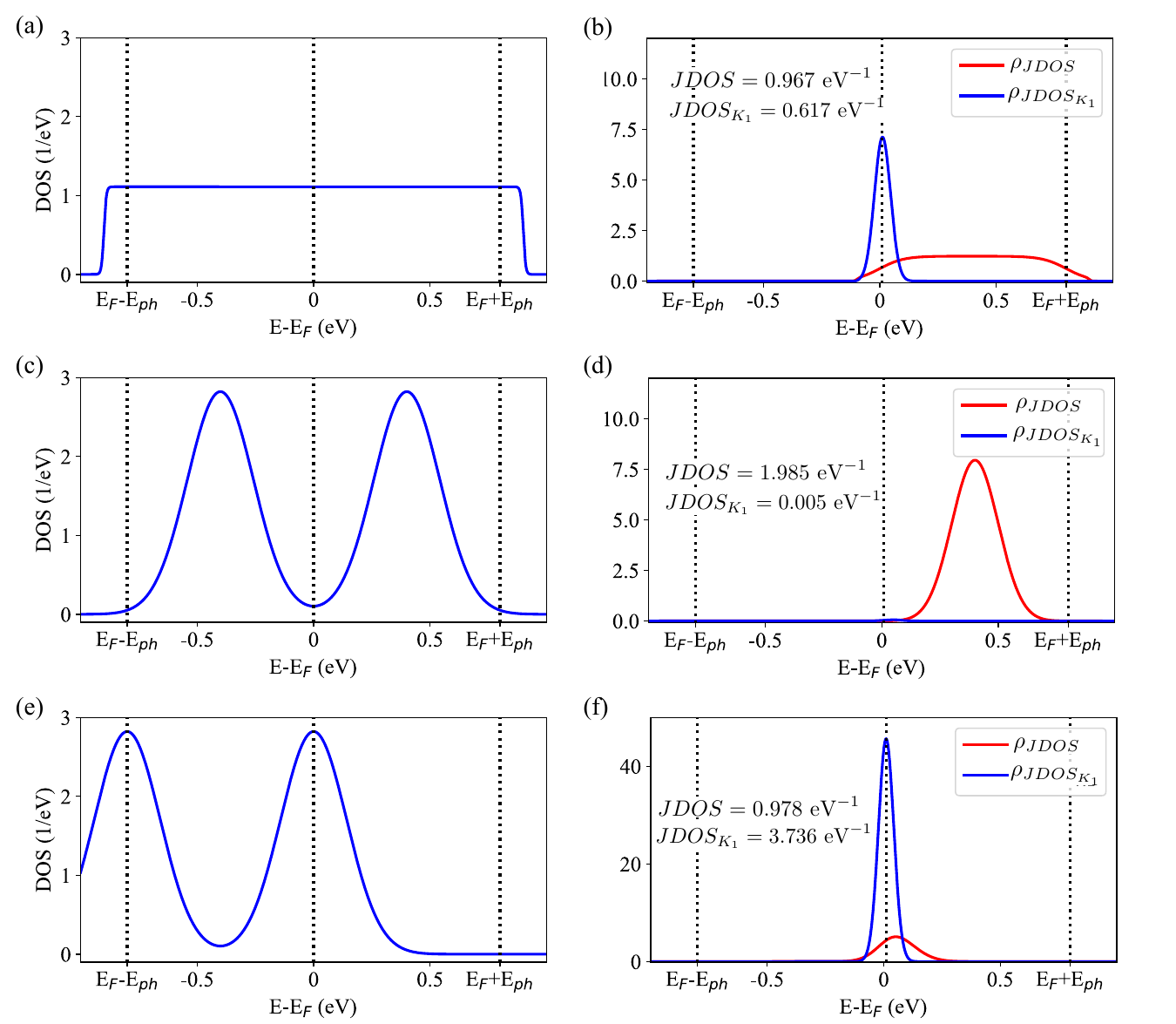}
\caption{{(a), (c) and (e) Show illustrative cases of DOS. (b), (d) and (f) demonstrate how $\rho_{JDOS}$ and $\rho_{JDOS_{K_1}}$ varies differs from each other for different energy distributions of states. We fixed total number of electrons to 2 and temperature, $T= 600$ K. Note the different vertical scale in panels (b), (d) and (f).}}
\label{fig:schem}
\end{figure}

Given the importance of indirect transitions, it is the densities of states at the energies corresponding to the initial and final states that determine the likelihood of exciting free carriers with photons in disordered alloys. This is why $JDOS$ was used  as an effective testing tool in such systems~\cite{Munday_2015,Munday_2020}.
In computing the joint density of states we assume that the transition is allowed if the initial state is occupied and the final state is empty. Denoting the density of states by $n(E)$, and noting that the probability of the state at energy 
$E-E_{ph}$ being occupied is 
$f(E-E_{ph},T)$ while the probability that the state at $E$ into which the electron is photoexcited being empty 
is $(1-f(E,T))$, where $f(E,T)$ is the Fermi function, the probability of exciting such a carrier is proportional to
\begin{equation}
    \rho_{JDOS}(E,T)=n(E-E_{ph})n(E)f(E-E_{ph},T)(1-f(E,T))\,.
\end{equation}
Assuming that the matrix element for the transition is not strongly dependent on energy in the relevant energy range, we use total $JDOS$ \cite{Munday_2015} as the measure of the number of excited electrons for a given photon energy,  
\begin{eqnarray}
JDOS(E_{ph}) = \int\ \rho_{JDOS}(E,T) dE \nonumber \\ \equiv\int {n(E-E_{ph})n(E)f(E-E_{ph},T)(1-f(E,T))} dE.
\label{JDOS}
\end{eqnarray}

Since the photon energies are of the order of eV, many electrons are photoexcited in the states far above the Fermi energy. { However, it is likely that inelastic electron-electron scattering, 
and other fast processes, will lead to recombination of high-energy electrons and holes. 
Thus, the higher energy electronic excitations are short-lived and will not significantly contribute to photoconduction at low bias. 
Therefore, we propose 
to introduce a kernel, $K$, in the definition of $JDOS$ to obtain a figure of merit (FOM) that emphasizes low energy excitations near $E_F$,}
\begin{eqnarray}
\fl FOM=JDOS_K(E_{ph})  = \int\ \rho_{JDOS_K}(E,T) dE \\
\nonumber \fl \qquad =\int{n(E-E_{ph})n(E)} f(E-E_{ph},T)(1-f(E,T)) K(E-E_F,T)dE\,.
\label{eq:JDOS_K}
\end{eqnarray}
Below we use and compare two (normalized) choices for the kernel $K(E-E_F,T)$, the thermally smeared Gaussian ($K_1$) and the step function ($K_2$)
\begin{eqnarray}
     K_1(E-E_F,T) = \frac{e^{-((E-E_F)/k_BT)^2}}{k_BT\sqrt{\pi}}, \\ 
     K_2(E-E_F,T) = 
     \cases
     {\frac{1}{2k_B T} \mbox{,for} -k_BT<E-E_F<k_BT\\
     0 \mbox{, otherwise}}
     \label{eq:kernel}
\end{eqnarray}
Regardless of the choice of kernel, the trends we find for the alloys studied do not change, as discussed below in section \ref{sec:results}. 

{Introduction of the kernel and emphasis on low energy excitations may result in very different trends exhibited by $JDOS_K$ compared to $JDOS$ depending on the density of states in the material. In Fig. \ref{fig:schem}, we illustrate this using three cases: (1) constant DOS between $E_F-E_{ph}$ to $E_F+E_{ph}$ results in $JDOS$ and $JDOS_{K_1}$ that are of the same order, see Figs. \ref{fig:schem} (a-b), (2) peaks in the DOS separated by the photon energy, $E_{ph}$, that are located away from the Fermi surface result in $JDOS_{K_1} \ll JDOS$, see Figs. \ref{fig:schem} (c-d), and (3) peaks in DOS separated by $E_{ph}$, with one peak near the Fermi energy result in $JDOS{K_1} \gg JDOS$, see Figs. \ref{fig:schem} (e-f). Importantly, for the latter two cases, that only differ by the shift of the DOS in energy, the $JDOS$ is comparable, but the $JDOS_K$ values are dramatically different. This highlights importance of appropriate choice of figure.}

\section{Computational Methods}
The simulation results were based on the electronic-structure method of Korringa, Kohn, and Rostoker (KKR) \cite{Stocks_94,Stocks_92,Soven_67,Korringa_47,Kohn_54} in combination with the Coherent-Potential Approximation (CPA) \cite{Stocks_94,Stocks_92,Soven_67} which handles both electronic structure and disorder (chemical or magnetic) on equal footing. The multi-atom KKR-CPA software was originally developed at Oak Ridge National Laboratory/University of Cincinnati  \cite{KKR_code}, to determine the electronic structure of complex alloys. The version used for this investigation is the full potential KKR-CPA developed by the J{\"u}lich group \cite{Russmann_2022,Schmitt_2022}. For these calculations, the full potential and charge density were determined  using a maximum angular momentum of $\ell_{max}$=4 for the spherical harmonic basis, and the Vosko, Wilks \& Nussar (VWN) local density functional \cite{Vosko_80,Wilk_1980} was employed for determining the exchange-correlation potential and energy.  The  $\vec{k}$-space integration is performed using the special $\vec{k}$-point integration technique~\cite{Chadi_73,Baldereschi_73,Monkhorst_76}.  A 16 × 16 × 16 $\vec{k}$-mesh is used. The rectangular energy contour contained 28 energy points.  The ground-state volumes were determined by calculating the total energy for 9 different lattice constants and fitting to the Birch– Murnaghan equation of state to obtain the ground-state volume \cite{Birch_47,Murnaghan_37}. After obtaining the ground-state volume, the electronic DOS was calculated using the Green’s function on an energy grid parallel to the real energy axis but with a shift of 0.0025 Ryd. in the complex energy plane.

\begin{figure}
\centering
\includegraphics[width=\columnwidth]{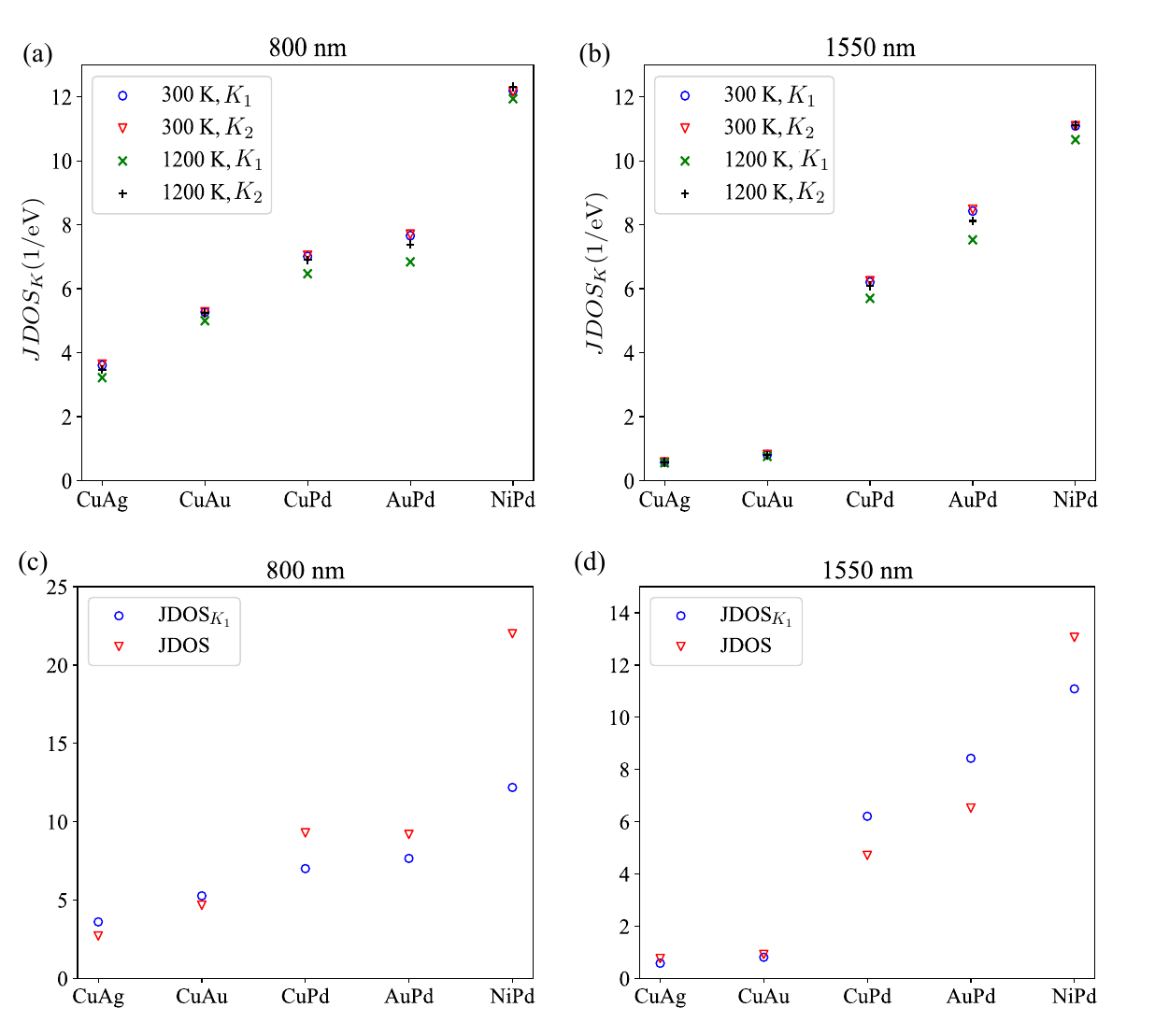}
\caption{(a),(b) $JDOS_{K_1}$ and $JDOS_{K_2}$ defined in Eqs. \ref{eq:JDOS_K} and \ref{eq:kernel} at $300 K$ and $1200 K$ for different transition-noble metal alloys with 50-50 ratio using (a) 800 nm and (b) 1550 nm light. (c),(d) Comparison of $JDOS$ and $JDOS_{K_1}$ for (c) 800 nm and (d) 1550 nm light shows quantitatively different trends.}
\label{fig:JDOS_K}
\end{figure}

\begin{figure}
\centering
\includegraphics[width=0.8\columnwidth]{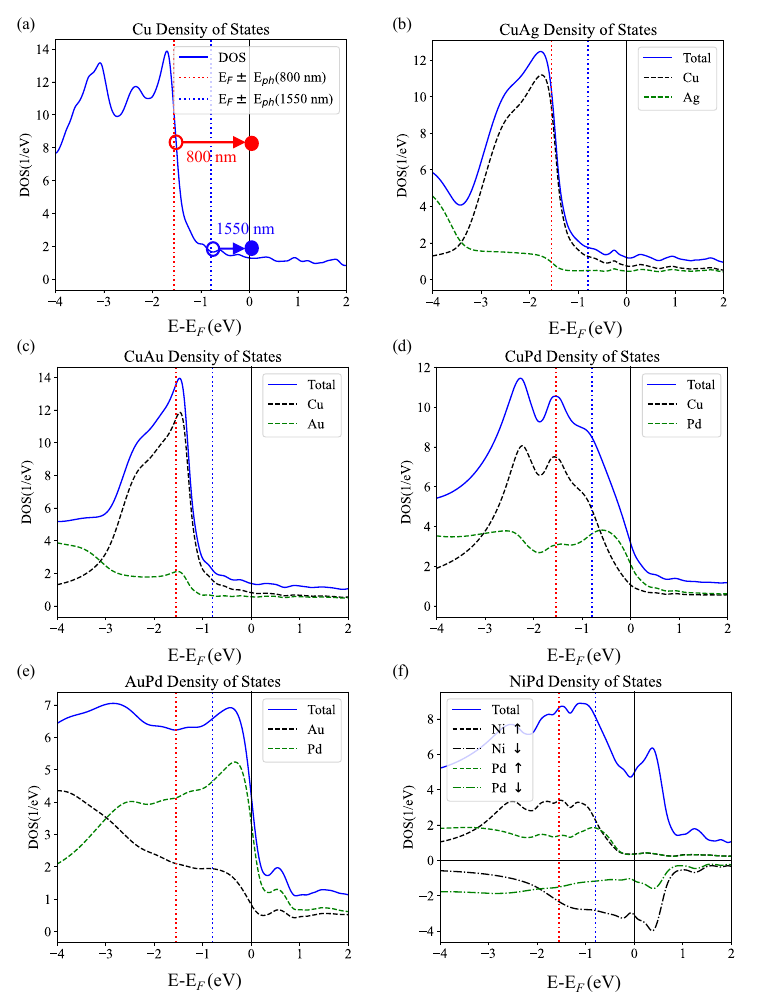}
\caption{The figure shows density of states for (a) Cu, (b) Cu$_{50}$Ag$_{50}$, (c) Cu$_{50}$Au$_{50}$, (d) Cu$_{50}$Pd$_{50}$, (e) Au$_{50}$Pd$_{50}$ and (f) Ni$_{50}$Pd$_{50}$. (a) also shows excitation of electron (solid circle) to around $E_F$ after optical excitation leaving a hole (empty circle) behind, using 800 nm (red) and 1550 nm light (blue).}
\label{fig:dos}
\end{figure}


\section{\label{sec:results}Results and Discussion}


$JDOS_{K}$ for disordered alloys Cu$_{50}$Ag$_{50}$, Cu$_{50}$Au$_{50}$, Cu$_{50}$Pd$_{50}$, Au$_{50}$Pd$_{50}$ and Ni$_{50}$Pd$_{50}$ is shown in Fig. \ref{fig:JDOS_K} for the two choices of the kernel and two different temperatures. For a particular choice of alloy, the variation in $JDOS_K$ for the two different kernels is small. The maximum difference is for Au$_{50}$Pd$_{50}$ at 1200 K and is still less than 8$\%$ for both 800 nm and 1550 nm wavelength, see Table \ref{table_K}. In addition, the normalized kernels result in almost temperature independent values for $JDOS_{K}$, see Fig \ref{fig:JDOS_K}. Thus, the trend in $JDOS_{K}$ values remains unchanged as components are varied, regardless of chosen kernel or temperature. 
Ni$_{50}$Pd$_{50}$ generates the most hot carriers according to our calculation. Below we compare and contrast the alloys considered.

Transition metals that have almost fully filled $d$-orbitals are well-suited for efficient hot carrier generation \cite{AuPd_AM,Fatti_2000,CuPd_aip}. Optical transition in an isolated atom usually require parity change up to leading order due to electric dipole term, $\propto \braket{\psi_f|\hat{\bm{r}}|\psi_i}$, thus $d\rightarrow d$ orbital transitions are forbidden. However, for disordered solids, broken inversion symmetry allows mixing of states with different parity. In addition, indirect transitions probability $\propto \int d \bm{r} e^{i(\bm{k}_i-\bm{k}_f).\bm{r}} \braket{u_{\bm{k}_f}|{\bm{r}}|u_{\bm{k}_i}}$ is generally non-zero due to disorder-assisted momentum change, see Sec. \ref{sec:indirect}. Thus, $d\rightarrow d$ band transitions are allowed for disordered alloys.




Low bias photoconductance depends on excitations near the Fermi surface, due to small lifetimes of electrons far from the Fermi surface. Thus, important excitations are ones where electrons are excited from around $E_F-E_{ph}$ to empty states near $E_F$. Thus, ideally we would want a tall and sharp peak in DOS at $E_F-E_{ph}$, indicating large number of filled states, and another peak of equal height at $E_F$, indicating the same number of empty states. In real materials this may not be possible. Below we use information about DOS peaks in pure metals, alloy them with carefully chosen components to align these peaks with $E_F-E_{ph}$. This is done by choosing alloying components with appropriate electronegativities that result in desired charge transfer between the components and, thus, desired change in position of the DOS peaks.

\begin{figure}
\centering
\includegraphics[width=0.8\columnwidth]{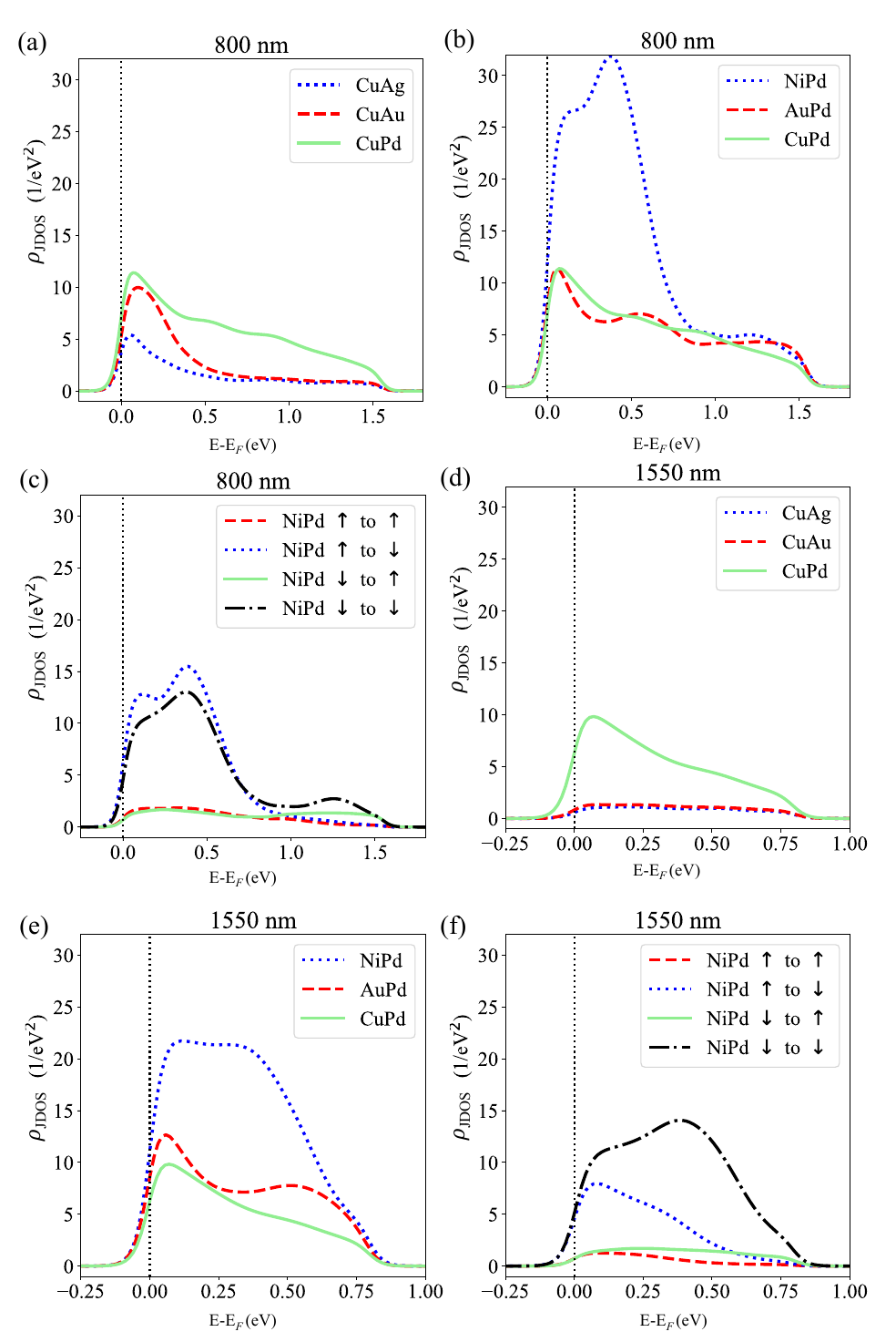}
\caption{The figure shows $\rho_{JDOS}$ for (a),(d) Cu alloys with 50-50 ratio, (b),(e) Pd alloys with 50-50 ratio, and (c),(f) Spin-split Ni$_{50}$Pd$_{50}$ for 800 nm (a-c) and 1550 nm  (d-f) wavelengths at $T = 300$ K.}
\label{fig:jdos}
\end{figure}

In its atomic form, Cu has electronic configuration $3d^{10}4s^{1}$. In solid form, this translates to filled $3d$ bands below $E_F$. Since $3d$ orbitals are closer to the nucleus than their  $4s$ counterparts which have comparable energy, overlap between $d$-orbitals of two neighboring Cu atoms is small, and so the $3d$-bands are narrower than 4$s$ band. The peak of the Cu DOS, arising largely from these narrow $3d$ bands, is near but not exactly at $E_F-E_{ph}$(800 nm), see Fig.~\ref{fig:dos}(a). As we will see below, alloying can potentially help align the peak with it. However, first, we pick a baseline disordered alloy. {Ag, which is a period below Cu. 
 has a Pauling electronegativity very close to that of Cu (1.93 and 1.90 for Ag and Cu, respectively) \cite{pauling1960}. Thus, there is only a small amount charge transfer between Cu and Ag on alloying. However, Ag bands are buried deep below $E_F$, see Fig.~\ref{fig:dos}(b). 
 Thus, the position of the Cu peak is virtually unchanged and Cu$_{50}$Ag$_{50}$ is a good baseline alloy.} 
$\rho_{JDOS}$ at 300 K also has a peak just above $E_F$, showing that most states are excited to just above $E_F$ when 800 nm light is used, see Fig.~\ref{fig:jdos}(a).
For 1550 nm, $\rho_{JDOS}$ has a flatter energy distribution, which comes from flat DOS near $E_F-E_{ph}$(1550 nm), see Figs.~\ref{fig:jdos}(d) and \ref{fig:dos}(b) respectively. We use $\rho_{JDOS}$ in Fig. \ref{fig:jdos} to calculate our figure of merit, $JDOS_K$. The kernel used for $T = 300 K$ has a width $~2k_BT \approx 0.05$ eV, thus only states very close to $E_F$ from $\rho_{JDOS}$ are captured. 

To align Cu $d$-band peak with $E_F-E_{ph}$(800 nm), 
{we need to move $E_F$ into the $d$-bands by removing electrons which can be done by alloying with groups to the left of Cu, Ag and Au in the periodic table.}
By comparing Figs. \ref{fig:dos}(b) and \ref{fig:dos}(c), we see that peak in Cu$_{50}$Au$_{50}$ is better aligned with $E_F-E_{ph}$(800 nm) than for Cu$_{50}$Ag$_{50}$. $\rho_{JDOS}$ for 800 nm has a peak similar to Cu$_{50}$Ag$_{50}$, but the states that were at much higher energy are now closer to $E_F$. This boosts $\rho_{JDOS}$ near $E_F$, see Fig. \ref{fig:jdos}(a), leading to increase in $JDOS_{K_1}$ for Cu$_{50}$Au$_{50}$ because states closer to $E_F$ have higher weight due to longer lifetimes, see Fig. \ref{fig:JDOS_K}.


Note the stark difference in values of $JDOS_{K_1}$ for 800 nm and 1550 nm, for both Cu$_{50}$Ag$_{50}$ and Cu$_{50}$Au$_{50}$ in Fig. \ref{fig:JDOS_K}. This is due to fewer states near $E_F-E_{ph}$(1550 nm) when compared to $E_F-E_{ph}$(800 nm)
, see Figs. \ref{fig:dos}(b) and \ref{fig:dos}(c). Even an increase in ratio of Au is unlikely to boost states near $E_F-E_{ph}$(1550 nm), as increase from Cu DOS peak moving closer to $E_F-E_{ph}$(1550 nm) will be negated by a corresponding reduction in peak height. Hence, a different component should be chosen to accomplish this. Pd being in the same period as Ag but with one less valence electron, has its 4$d$ orbitals filled with 9 electrons. This provides a broad DOS of width $\sim$4 eV that is pinned to $E_F$. Pd being more electronegative than Cu takes some charge away, pushing Cu states nearer to $E_F-E_{ph}$(1550 nm), see Fig. \ref{fig:dos}(d). Additionally, there is strong mixing between Cu and Pd resulting in a boost of occupied states near $E_F-E_{ph}$(1550 nm). 
As a result $\rho_{JDOS}$ for Cu$_{50}$Pd$_{50}$ has much higher peak than Cu$_{50}$Ag$_{50}$ and Cu$_{50}$Au$_{50}$, see Fig. \ref{fig:jdos}(d). This substantially increases its $JDOS_K$ for 1550 nm compared to Cu$_{50}$Ag$_{50}$, see Fig. \ref{fig:JDOS_K}(b) and Table \ref{table_jdos}. Transfer of negative charge from Cu moves the $d$-states peak nearer to $E_F - E_{ph}$(800 nm) but peak height is reduced from hybridization. However, empty states above $E_F$ increase. The overall effect is a substantial increase in $JDOS_K$ for 800 nm, see Fig. \ref{fig:JDOS_K}(a). 



\begin{table}[h]
\centering
\begin{tabular}{ |p{1.5cm}|p{1.5cm}|p{1.5cm}|p{1.5cm}|p{1.5cm}|}
\hline 
& 800 nm & 800 nm & 1550 nm & 1550 nm\\ 
\hline
 Alloy& {$JDOS$ (1/eV)} & {$JDOS_{K_1}$ (1/eV)} & {$JDOS$ (1/eV)} & {$JDOS_{K_1}$ (1/eV)}\\
 \hline
 \hline
 Cu$_{50}$Ag$_{50}$ & 2.71 & 3.61 & 0.76 & 0.58 \\
 Cu$_{50}$Au$_{50}$ & 4.67 & 5.27 & 0.92 & 0.81\\
 Cu$_{50}$Pd$_{50}$ & 9.29 & 7.01 & 4.71 & 6.21\\
 Au$_{50}$Pd$_{50}$ & 9.19 & 7.66 & 6.52 & 8.43\\
 Ni$_{50}$Pd$_{50}$ & 22.00 & 12.19 & 13.06 & 11.09\\
 \hline
\end{tabular}
 \caption{Comparison of $JDOS$ and $JDOS_{K_1}$ for NIR  wavelengths 800 nm and 1550 nm at $300$ K. Note the increase in values of figure of merit down the table, $JDOS_{K_1}$ as we strategically switch alloy components. The trend can be {quantitatively} different for $JDOS$ and the proposed figure of merit, $JDOS_{K}$ (for kernel $K_1$).}
 \label{table_jdos}
\end{table}

So far we have optimized for the number of occupied states near $E_F-E_{ph}$, now we focus on boosting empty states near $E_F$. Au being only slightly more electronegative than Pd will attract some negative charge to itself pushing some of Pd broad band states near the Fermi surface above $E_F$, see Fig. \ref{fig:dos}(e). Pd and Au both have broad bands composed of $4d$ and $5d$ that {mix} to give high DOS near $E_F-E_{ph}$(1550 nm). 
This leads to increase in $\rho_{JDOS}$ near fermi surface compared to Cu$_{50}$Pd$_{50}$, see Fig. \ref{fig:jdos}(e). Consequently, $JDOS_{K_1}$ increases compared to Cu$_{50}$Pd$_{50}$, see Fig. \ref{fig:JDOS_K}(b). For 800 nm increase in empty states above $E_F$ more than compensates for the relative decrease in filled states below $E_F$, resulting in slight gain in $JDOS_{K_1}$ over Cu$_{50}$Pd$_{50}$, see Fig. \ref{fig:JDOS_K}(a).




Another way to boost states both below and above $E_F$ is to use a ferromagnetic components for alloying. Ferromagnetism splits bands, and thus DOS, pushing a few states above and few states below $E_F$. Ni$_{50}$Pd$_{50}$ is ferromagnetic as Ni magnetizes Pd leading to a split in DOS, see Fig.~\ref{fig:dos}(f). From Table \ref{table_K}, we note $\uparrow$ to $\downarrow$ and $\downarrow$ to $\downarrow$ are dominant contributions to its $JDOS_K$. In presence of spin-orbit interaction, a photon absorption could result in electron spin-flip or spin-preserving transitions because spin and spatial degrees of freedom are coupled. 
Thus, we add both spin-flip and spin-preserving transitions to calculate $JDOS_K$ for Ni$_{50}$Pd$_{50}$.

{We consider only binary alloys with a 50-50 composition.
This is largely to give a clear demonstration of the scheme. Note that the peak in DOS above $E_F$ lies near 0.5 eV, see Fig. \ref{fig:dos}(f). Thus, there is room for improvement in $JDOS_K$ by decreasing Ni composition in Ni$_{50}$Pd$_{50}$. However, we leave exploration of different ratios for future work.}

\begin{table}[h]
\centering
\begin{tabular}{ |p{1.5cm}||p{1.2cm}|p{1.2cm}||p{1.2cm}|p{1.2cm}| }
 \hline
 & \multicolumn{2}{|c|}{$T = 300 K$ } &\multicolumn{2}{|c|}{$T = 1200 K$ } \\
 \hline
 Alloy & 800 nm & 1550 nm & 800 nm & 1550 nm \\
 \hline
 Cu$_{50}$Ag$_{50}$ & 3.61 (3.64) & 0.58 (0.58) & 3.22 (3.47)& 0.56 (0.58)\\
 Cu$_{50}$Au$_{50}$ & 5.27 (5.27) & 0.81 (0.81) & 5.00 (5.24)& 0.76 (0.80)\\
 Cu$_{50}$Pd$_{50}$ & 7.01 (7.04) & 6.21 (6.24) & 6.47 (6.90)& 5.70 (6.09)\\
 Au$_{50}$Pd$_{50}$ & 7.66 (7.70) & 8.43 (8.48) & 6.84 (7.38) & 7.53 (8.12)\\
 Ni$_{50}$Pd$_{50}$ & 12.19 (12.16) & 11.09 (11.09) & 11.94 (12.31) & 10.66 (11.11)\\
 ($\uparrow$ to $\uparrow$) & 0.86 (0.86) & 0.72 (0.72) & 0.83 (0.85)& 0.67 (0.71)\\
($\uparrow$ to $\downarrow$) & 6.09 (6.08) & 4.54 (4.54) & 5.92 (6.13)& 4.23 (4.48)\\
 ($\downarrow$ to $\downarrow$) & 4.55 (4.53) & 5.12 (5.10) & 4.52 (4.63)& 5.05 (5.20)\\
($\downarrow$ to $\uparrow$) & 0.69 (0.69) & 0.72 (0.72) & 0.68 (0.69)& 0.70 (0.72)\\
 \hline
\end{tabular}
 \caption{$JDOS_{K_1}$ ($JDOS_{K_2}$) for kernels $K_1$ and $K_2$ defined in Eq.~\ref{eq:kernel}, for transition-noble metal alloys using NIR  wavelengths 800 nm and 1550 nm and temperatures $300$ K and $1200$ K.}
 \label{table_K}
\end{table}

Comparing $JDOS$ with $JDOS_K$ in Table \ref{table_jdos} we note that $JDOS$ follows a similar trend. Thus, had we used $JDOS$ without the kernel, we would have been led to the same conclusion that Ni$_{50}$Pd$_{50}$ outperforms the other alloys. However, careful consideration of the numbers reveals that degree of improvement as one goes down the table is not the same for $JDOS$ and $JDOS_K$. In Table \ref{table_jdos}, the improvement in hot carrier generation using 800 nm light on switching Au with Pd in the Cu alloy is $98\%$, while it was only $33\%$ from $JDOS_{K_1}$. 
$JDOS$ also significantly overestimates improvement in switching Au with Ni in Pd alloy, compared to estimates from $JDOS_{K_1}$. This is largely because highest point in DOS of Ni$_{50}$Pd$_{50}$ lies $\sim 0.5$ eV away from the Fermi surface, see Fig. \ref{fig:jdos}(b). $JDOS_{K_1}$ down-weights these states as they will not contribute to photoconduction, while $JDOS$ includes them and hence the overestimates improvement in hot carrier generation. Additionally, $JDOS$ and $JDOS_{K_1}$ suggest opposite trend for Cu$_{50}$Pd$_{50}$ and Au$_{50}$Pd$_{50}$. Thus, similarity of the trend between $JDOS$ with $JDOS_K$ is largely due to our careful choice of components which for the most part have $\rho_{JDOS}$ peaks near the Fermi surface, see Fig. \ref{fig:jdos}(a-f). { However, for a different set of alloys, $JDOS$ and $JDOS_K$ can suggest different trends in hot generation depending on the energy distribution of states, as illustrated in section 2.}

\section{Conclusion}

{ In summary, above we provided a detailed justification for  use of $JDOS$-like quantities to rank the performance of alloys for generation of photo-induced hot carriers. We explicitly demonstrated that the linear response contributions of disorder-assisted (indirect) transitions and direct transitions  are of the same order. Furthermore, we proposed a new figure of merit, $JDOS_K$, that is appropriate for low-bias photoconduction. We implemented our methodology, in conjunction with ab initio KKR-CPA method that we used to calculate the density of states, to rank several disordered noble-transition metal alloys for low-bias photoconduction. }



{ 
Within this series we showed that for a 50-50 ratio Ni$_{50}$Pd$_{50}$ is likely to have the highest hot-carrier generation. This is due to ferromagnetic split of DOS contributions from both Ni and Pd, which boosts filled states above and empty states near the Fermi surface.}

{ We expect that our method is generally applicable to a wide range of alloys. It is possible that in some alloys the main source of scattering is not from the Coulomb potential, for example, long range strain field (and hence small-momentum scattering). In such a case, the contribution of indirect transitions may be lower than our estimate, even though local breaking of point group symmetries by the strain still enables non-momentum conserving photoexcitations. We do not expect these cases to be common among the $d$-electron systems. Secondly, it is conceivable that for sufficiently dilute alloys that rare realizations of disorder give dominant contribution to the photoexcitations, substantially modifying the mean field KKR-CPA results (providing essentially an analog of Griffiths phases in magnets). It seems to us, that this would likely occur only in the strongly interacting electron limit, less relevant to the noble metal alloys, but further investigations of this possibility are needed. Away from the dilute limit we expect the methods presented here to give a reliable tool for ranking alloys for low-bias photoconduction.}





\section*{Acknowledgements}

E.T., K.M.M., P.T.S. and W.A.S. acknowledge funding from the National Science Foundation under research Award No. DMR-2114304.

\section*{References}


\end{document}